\documentclass[pdflatex,sn-nature]{sn-jnl}

\usepackage{graphicx}%
\usepackage{multirow}%
\usepackage{amsmath,amssymb,amsfonts}%
\usepackage{amsthm}%
\usepackage{mathrsfs}%
\usepackage[title]{appendix}%
\usepackage{xcolor}%
\usepackage{textcomp}%
\usepackage{manyfoot}%
\usepackage{booktabs}%
\usepackage{placeins}

\theoremstyle{thmstyleone}%
\theoremstyle{thmstyletwo}%

\theoremstyle{thmstylethree}%

\begin{document}
\fontsize{11pt}{13.75pt}\selectfont
\newgeometry{
    left=1in,
    right=1in,
    top=1in,
    bottom=1in,
    bindingoffset=0mm
}

\title[Raman-detected quantum dot microscopy]{Raman-detected quantum dot microscopy for nanoscale electrostatic potential imaging}


\author*[1,2]{\fnm{Ji\v{r}\'i} \sur{Dole\v{z}al}}\email{jiri.dolezal@uochb.cas.cz}

\author[2,3]{\fnm{Amandeep} \sur{Sagwal}}

\author[1]{\fnm{Rodrigo Cezar} \sur{de Campos Ferreira}}

\author*[1,2]{\fnm{Martin} \sur{\v{S}vec}}\email{svec@fzu.cz}

\affil*[1]{\orgdiv{Institute of Organic Chemistry and Biochemistry}, \orgname{Czech Academy of Sciences}, \orgaddress{\street{Flemingovo náměstí 542/2}, \postcode{CZ16000}, \state{Praha 6}, \country{Czech Republic}}}

\affil[2]{\orgdiv{Institute of Physics}, \orgname{Czech Academy of Sciences}, \orgaddress{\street{Cukrovarnická 10/112}, \postcode{CZ16200}, \state{Praha 6}, \country{Czech Republic}}}

\affil[3]{\orgdiv{Faculty of Mathematics and Physics}, \orgname{Charles University}, \orgaddress{\street{Ke Karlovu 3}, \postcode{CZ12116}, \state{Praha 2}, \country{Czech Republic}}}


\abstract{Quantification of electrostatic potentials at the nanoscale is crucial for understanding the principles governing properties of materials across multiple length scales. Currently, one of the most successful approaches relies on the charging response of a molecular quantum dot, suspended on a tip of a scanning probe microscope and measured using dynamic force spectroscopy. We investigate the possibility of an optical detection, aiming to improve the speed and reduce the complexity of this measurement scheme. We show that the integrated tip-enhanced Raman scattering intensity strongly correlates with the charge state of the quantum dot, and use it to map the electrostatic potential of a single atom. A quantitative equivalence with the established force spectroscopy method is found. We address the underlying photophysical principle of this new method by measuring the Raman spectra as a function of excitation wavelength and the molecular quantum dot charge. We reveal that the observed Raman intensity variations are primarily driven by transitions between resonant and non-resonant Raman scattering regimes of the molecule.
}




\maketitle

\section*{Introduction}\label{sec1}

Optical probes capable of sensing local fields at the nanometer scale are central to modern nanophotonics \cite{Novotny2006}, as they would enable access to structural, electronic, magnetic, and chemical properties of nanostructures beyond the diffraction limit of light. Quantum dots (QDs) are of particular interest for such precise nanoscopic sensing \cite{Michaelis2000-qn,Cadeddu2016-wp,Bian2021-lt} due to their discrete energy spectra, strong light-matter interaction, and their tunability by externally applied fields. Among the prominent examples are the nitrogen--vacancy (NV) \cite{Maze2008-pw} centers in diamond, employed in atomic force microscopy as a high-precision magnetometer. However, the finite size of the QD host crystal typically limits the spatial resolution to tens of nanometers. Conversely, sub-nanometer studies are routinely performed with organic-molecule QDs, used as a functionalization of atomically sharp tips in cryogenic scanning probe microscopy. For example, functionalization with carbon monoxide (CO) enables bond-resolved imaging \cite{Fatayer2019-hx,Chiang2014-az}, nickelocene provides spin sensitivity \cite{Verlhac2019-lk,Czap2019-rg}, and some rylene derivatives such as perylene- and napthalo tetracarboxylic dianhydrides (PTCDA and NTCDA) were shown to be convenient as electrostatic \cite{Wagner2015-yo} and magnetic \cite{Esat2024-ni} probes. Despite such versatility, these probes rely on detecting force gradients as a frequency shift in dynamic atomic force microscopy (AFM) or as inelastic electron tunneling in STM, both of which have bandwidths fundamentally limited by the complexity of their measurement schemes. Increasing the efficiency and improving the response time would require an alternative approach based on a different quantity for the detection of signals relevant to the particular probe response to the measured field.

Recently, a purely optical readout of signals from nanoscale probes has been proposed and demonstrated as a viable method for electrostatic nanoscale mapping. The principles employed to achieve this were measurements of electrically induced Stark shifts of the stretching mode in a CO-functionalized STM tip via tip-enhanced Raman spectroscopy (TERS) \cite{Lee2018-ij}, and detection of Stark shifts of electronic transitions of a PTCDA anion by STM-induced luminescence (STML) \cite{Friedrich2024-zp}. Detection by light holds the promise of faster readout and a higher signal-to-noise ratio. However, neither of these methods has yet achieved quantitative and reliable nanoscale electrostatic mapping comparable to Kelvin probe force microscopy (KPFM) \cite{Sadewasser2018-hw} or scanning quantum dot microscopy (SQDM) \cite{Wagner2015-yo}, which are well established in the field of dynamic atomic AFM.

Here, we build upon recent advances in TERS \cite{Zhang2013-mj,Jaculbia2020-tz,Lee2019-nh}, specifically the demonstrations of strong chemical enhancement that reportedly occurs upon contacting a molecule with the metal tip \cite{Cirera2022-wl,Liu2023-hh,Yang2023-ot} and of the resonant enhancement achieved when the laser energy matches specific molecular excited states \cite{Jaculbia2020-tz,De_Campos_Ferreira2024-jb}. We exploit these phenomena in order to track the charging of a tip-attached PTCDA molecule in real-time to reconstruct the distribution of the local electrostatic potential. We corroborate the equivalence of our optical method to the AFM-based SQDM through a correlation between the abrupt changes detected in the overall TERS intensity and the charging thresholds detected at the same time via the frequency shift in AFM. These direct simultaneous TERS and AFM measurements of the electrostatic fingerprint of a subsurface defect provide a direct comparison of the yields and accuracy of the two methods. Finally, we devise a fast feedback-driven scheme for quantitative mapping of the electrostatic potential and apply it to a surface adatom.

\section*{Main}\label{sec2}

\subsection*{TERS readout of QD charge state}

\begin{figure*}[t]
\centering
\includegraphics[width=\textwidth]{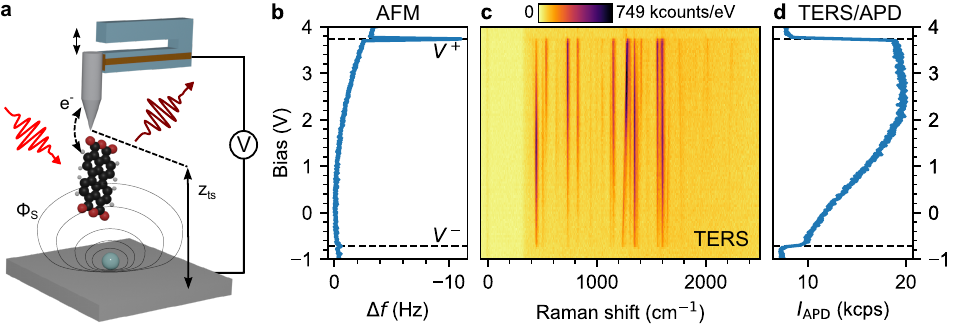}
\caption{\textbf{Optically detected SQDM experiment.}
\textbf{a,} Schematic of the TERS--SQDM experimental setup. A PTCDA quantum dot (QD) is attached to the tip apex. Changes in the QD charge state are detected either through the atomic-force response or through changes in the TERS intensity.
\textbf{b,} Frequency-shift, $\Delta f(V)$, dependence with two charging peaks ($V^{+}$, $V^{-}$) marked by dotted lines.
\textbf{c,} Heatmap of voltage-dependent Raman spectra acquired with a 1.96~eV laser.
\textbf{d,} Integrated photon intensity, $I_{\mathrm{APD}}$, above 640~nm ($\sim 156$~cm$^{-1}$) recorded with an APD during the voltage sweep shown in \textbf{b}. The intensity changes coincide with the charging peaks. All data were acquired with the same oscillating tip using an amplitude of 50~pm and a tip--sample distance of 2.7~nm. The laser power was 10~$\mu$W.}
\label{fig1}
\end{figure*}

First, to probe the capability of TERS for detecting the charge state of the single-molecule QD, we prepared an atomically sharp Ag-terminated tip attached to a qPlus sensor prong, with an intense nanocavity plasmon spectrally matching the HeNe laser excitation source (1.96~eV). A single PTCDA molecule was lifted from the Ag(111) surface by a controlled approach of the tip to one of the corner oxygen atoms of PTCDA at 1~mV bias, followed by retraction to a tip-sample distance ($z_\mathrm{ts})$ of 2.7~nm (Extended Data Fig.~\ref{figE1}), as documented in previous works \cite{Wagner2015-yo,Friedrich2024-zp,Gong2025-gb}. At zero bias, the suspended molecule is known to adopt an anionic charge state \cite{Friedrich2024-zp,De_Campos_Ferreira2024-jb} that can be switched to neutral or dianionic state by applying negative or positive bias on the sample, respectively \cite{Wagner2015-yo}. A bias sweep (in Fig.~\ref{fig1}\textbf{b}) reveals the charge state switching levels manifested as telltale dips in $\Delta f(V)$ measured by AFM. These arise from the abrupt changes in the electrostatic tip--sample force when the charge state of the molecule switches during an oscillation cycle of the tip \cite{Wagner2015-yo}. Following the nomenclature of Wagner et al. \cite{Wagner2015-yo}, we denote the charging bias levels of this molecular QD as $V^{+}$ (anion $\leftrightarrow$ dianion) and $V^{-}$ (anion $\leftrightarrow$ neutral).
Simultaneously with the $\Delta f(V)$ measurement, we recorded TERS spectra and the integrated photon intensity $I_{\mathrm{APD}}(V)$ measured with a single-photon avalanche photodetector (APD) (Fig.~\ref{fig1}\textbf{c,d}). With the incident laser illuminating the tip apex, Raman scattering on the suspended PTCDA is induced, and we observe its strong dependence on the bias voltage applied to the sample. In particular, the overall Raman signal intensity and spectral envelope show a striking correlation with the regions of stable QD charge states. Between $V^{-}$ and $V^{+}$, where the molecule is in an anionic state, an intense Raman spectrum emerges (see Extended Data Fig.~\ref{figE3}), closely resembling the previously reported resonant Raman fingerprint of the PTCDA anion in a break junction \cite{De_Campos_Ferreira2024-jb}. We observe that above $V^{+}$, the Raman intensity drops by more than an order of magnitude as the molecule adopts the dianionic charge state (see detailed spectra in Extended Data Fig.~\ref{figE3}). Similarly, below $V^{-}$, the molecule is stabilized in the neutral charge state, yielding also a low TERS intensity, especially in the spectral region below 1200~cm$^{-1}$.

\begin{figure}[t]
\centering
\includegraphics[width=0.5\textwidth]{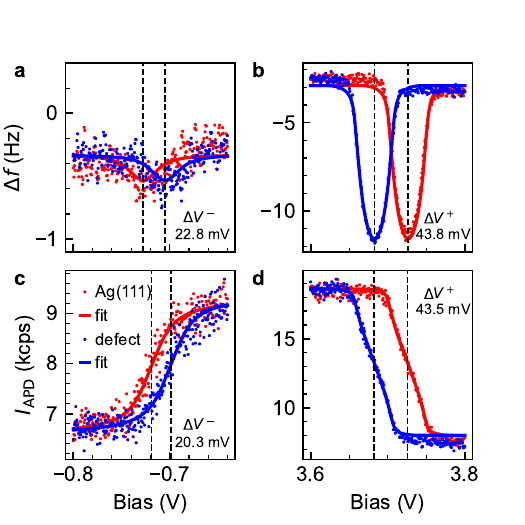}
\caption{\textbf{Equivalence of TERS- and force-detected QD spectroscopy.}
\textbf{a--d,} Shifts of the $V^{+}$ and $V^{-}$ thresholds, indicated by dashed lines, measured above a subsurface defect on Ag(111) (blue) and 7~nm away on the bare surface (red).
\textbf{a,b,} Frequency-shift, $\Delta f(V)$, signal.
\textbf{c,d,} Raman intensity, $I_{\mathrm{APD}}$. Solid lines show fits using the frequency-shift response equation (\ref{df_model}) and the occupation-broadened TERS intensity function equation (\ref{Iapd_model}), respectively. The broadening parameters were fixed from the
oscillation-amplitude dependence shown in Extended Data Fig. \ref{figE5}, leaving only the charging threshold
\(V_0\) as a free parameter. Measurements were performed at $z_{\mathrm{ts}} = 2.7$~nm with an oscillation amplitude of 50~pm.}
\label{fig2}
\end{figure}

Due to these sharp intensity variations, the integrated TERS response of the PTCDA at the tip is an ideal quantity for tracking the charge transitions with high precision and efficiency. The $I_{\mathrm{APD}}$ curve obtained with an avalanche photodiode in Fig.~\ref{fig1}\textbf{d} for photon wavelengths above 640~nm shows sharp steps marking the two charging biases $V^{+}$ and $V^{-}$, perfectly coinciding with the charging dips in the simultaneously acquired $\Delta f(V)$. Detailed $\Delta f(V)$ and TERS integral intensity curves around both charging thresholds above a subsurface defect and on the pristine Ag(111) surface (see Fig.~\ref{fig2}) show an excellent correspondence between the center positions and widths of the charging features, extracted from their fits (see Methods). Notably, the width of the $V^{+}$ feature is dominated by the tip oscillation amplitude shown in Extended Data Fig.~\ref{figE5}, whereas $V^{-}$ exhibits additional laser-induced broadening (see Extended Data Fig.~\ref{figE4}) and an upshift in voltage with respect to the unilluminated case, while remaining largely unaffected by increasing the oscillation amplitude up to 100~pm. We suggest that either photoassisted neutralization or photoinduced modification of the adsorption geometry of the molecule on the tip, previously observed for larger tunneling currents flowing through the QD \cite{green2016scanning}, could modify the alignment and the lifetime of the QD electronic states with respect to the Fermi level of the tip. A detailed understanding of this phenomenon would require laser-power- and excitation-energy-dependent measurements.

\subsection*{Quantitative electrostatic-potential mapping with TERS intensity}

Having established that TERS-SQDM and AFM-based SQDM yield equivalent charging-bias thresholds for determining the electrostatic potential, we can investigate TERS-SQDM capability, advantages, and constraints in quantitative electrostatic potential imaging. To this end, we prepared an Ag STM tip mounted in a standard STM holder without AFM functionality and used it to obtain the electrostatic potential map solely with TERS-SQDM of an Ag adatom on Ag(111), a known benchmark system for AFM-SQDM \cite{Wagner2015-yo,Bolat2024-fr,Wagner2019-bb}. Because the acquisition of the point spectra on a spatial grid (or, equivalently, a stack of constant-height images acquired at biases around $V^{+}$ and $V^{-}$, see Extended Data Fig.~\ref{figE5_5}) is time-consuming and sensitive to both vertical and lateral drift, it was crucial to implement a feedback loop to track the charging threshold voltages. A related approach was recently introduced for AFM-SQDM by Wagner and Maiworm et al. \cite{Maiworm2021-pd,Wagner2019-bb}. The feedback loop substantially reduces acquisition time while enabling direct recording of $V^{+}$ and $V^{-}$ maps without complex postprocessing.

\begin{figure*}[t]
\centering
\includegraphics[width=\textwidth]{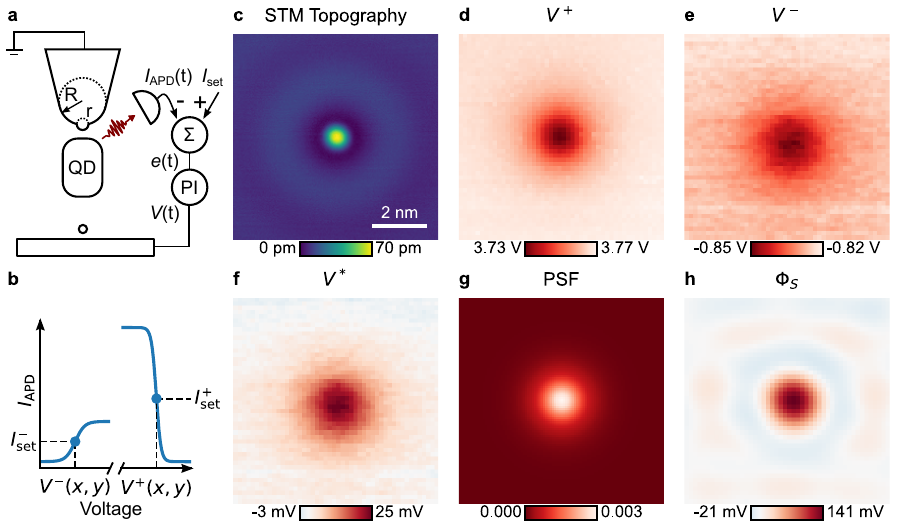}
\caption{\textbf{Electrostatic-potential imaging of an Ag adatom.}
\textbf{a,} Schematic of the feedback scheme used for TERS-SQDM. The STM tip is represented by a sphere-on-sphere geometry, consisting of a spherical body with radius of curvature ($R$), terminated by a smaller hemispherical apex of radius ($r$), 
following the tip model used to simulate sub-nanometre-resolved TEPL images\cite{Yang2023-ot}.
\textbf{b,} Schematic bias-dependent TERS intensity curves with the centers of the step used as setpoints to track $V^+$ and $V^-$.
\textbf{c,} STM topography of an Ag adatom on Ag(111) acquired in constant-current mode with an Ag tip at 100~mV and 10~pA.
\textbf{d,e,} $V^{+}$ and $V^{-}$ maps acquired with bias feedback at the tip--sample distance of $z_\mathrm{ts} = 2.6$~nm. Acquisition times were 3~min for the $V^{+}$ map and 30~min for the $V^{-}$ map.
\textbf{f,} $V^{*}$ map calculated from $V^{+}$ and $V^{-}$ using equation~(\ref{V_star}).
\textbf{g,} Normalized electrostatic point spread function calculated at the experimental imaging height for the sphere-on-sphere tip geometry with $R=50$ nm and $r=0.5$ nm.
\textbf{h,} Reconstructed surface electrostatic potential distribution, $\Phi_{\mathrm S}(x,y)$, obtained by regularized deconvolution of the $V^{*}$ map in \textbf{f} with the point spread function in \textbf{g}.} 
\label{fig3}
\end{figure*}

A characteristic difference and a major practical advantage of TERS-SQDM over AFM-SQDM is that the integrated Raman intensity is a monotonic function near $V^{+}$ and $V^{-}$, unlike the dip-like $\Delta f(V)$. Moreover, in the TERS-SQDM, there is no need for tip oscillation, which can yield sharper resolution of the $V^{+}$ charging threshold (see Extended Data Fig.~\ref{figE5}). We used a proportional-integral controller sketched in Fig.~\ref{fig3}\textbf{a} (see Methods for details) to maintain the $I_{\mathrm{APD}}$ setpoint at the center of the charging step illustrated in Fig.~\ref{fig3}\textbf{b} and acquired constant-height maps of $V^{+}(x,y)$ and $V^{-}(x,y)$. Both maps show a symmetrical perturbation centered on the adatom position identified in the STM topography (Fig.~\ref{fig3}\textbf{c--e}).

From $V^{+}$ and $V^{-}$, we calculated $V^*$ using equation~(\ref{V_star}), which represents the electrostatic potential $\Phi_S$ sensed by the QD at the imaging plane within the SQDM formalism \cite{Wagner2019-bb,wagner2019theory,Bolat2024-fr} (Fig.~\ref{fig3}\textbf{f}). Integration of $V^*$ over the imaging plane yields a dipole moment of 0.62~D, in excellent agreement with the previously reported AFM-SQDM value of 0.66~D for the same system \cite{Wagner2019-bb,Bolat2024-fr}. To reconstruct the quantitative surface potential $\Phi_\mathrm{S}(x,y)$ shown in Fig.~\ref{fig3}\textbf{h}, we deconvolved $V^*$ using a SQDM point spread function kernel \cite{Wagner2019-bb} shown in Fig.~\ref{fig3}\textbf{g} for a realistic sphere-on-sphere STM tip geometry (see Methods). This procedure converts the potential sensed at the QD position into a quantitative map of the electrostatic potential at the sample surface, thereby enabling spatially resolved electrostatic imaging. We identify a systematic error of TERS-SQDM arising from the upward shift of $V^-$ by 40 mV under illumination, which introduces a relative error in $V^*$ of less than $10^{-3}$ and is therefore negligible for the quantification of the surface dipole moment.

\subsection*{Tuning the laser in resonance with the charged state}

\begin{figure}[t]
\centering
\includegraphics[width=0.5\textwidth]{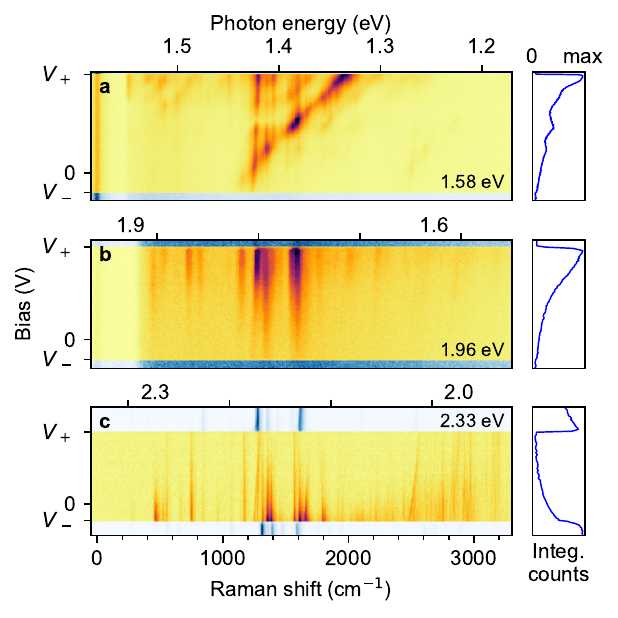}
\caption{\textbf{Voltage-dependent Raman spectra acquired with different laser energies.}
\textbf{a,} Raman spectra acquired with a 1.58~eV laser.
\textbf{b,} Raman spectra acquired with a 1.96~eV laser.
\textbf{c,} Raman spectra acquired with a 2.33~eV laser. The acquisition time per spectrum was 5~s in \textbf{a}, 5~s in \textbf{b}, and 3~s in \textbf{c}.}
\label{fig4}
\end{figure}

Finally, we turn our focus to the origin of the bias dependence of the TERS of the suspended PTCDA QD and its connection to its electronic transitions. TERS spectra and calculated integral intensity as a function of bias with three different laser excitation energies are shown in Fig.~4a--c. For excitation at 1.58~eV (Fig.~\ref{fig4}a) and 1.96~eV (Fig.~~\ref{fig4}b), the photon intensity profiles exhibit a similar behavior: the molecule in the anion state gives an intense TERS signal, in contrast to the dark neutral and dianion. 
\begin{table}[h]

\label{tab1}
\caption{Excited-state transition energies and dipole orientations for different charge states. TD-DFT values for the electronic transition energies are given for non-relaxed geometries; values corresponding to emission, calculated for relaxed geometries are provided in parentheses. Experimental data for $S_1^{2-}$ are presented in Extended Data Fig. \ref{figE6} and $S_2^{2-}$ were measured using Tip-enhanced photoluminescence (TEPL) of PTCDA on a MgO/Ag(100) decoupling surface.}
\begin{tabular}{@{}lllll@{}}
\toprule
\multirow{2}{*}{Charge state} &
\multirow{2}{*}{Excited state} &
\multicolumn{3}{c}{Transition energy (eV) to ground state} \\
\cmidrule(lr){3-5}
& & Experiment &
\begin{tabular}{@{}c@{}}TD-DFT\\Theory\end{tabular} &
\begin{tabular}{@{}c@{}}Transition dipole\\orientation relative\\ to molecular axes\end{tabular} \\
\midrule
neutral & $S_1$      & $2.47$ \cite{Kimura2019-yp}              & $2.91\,(2.24)$ \cite{De_Campos_Ferreira2024-jb} & long  \\
anion   & $D_1^{-}$  & $1.33$ \cite{Dolezal2022-ok}            & $1.69\,(1.34)$ \cite{De_Campos_Ferreira2024-jb} & long  \\
        & $D_2^{-}$  & $1.50$ \cite{Dolezal2022-ok}            & $1.95$ \cite{De_Campos_Ferreira2024-jb}         & short \\
        & $D_3^{-}$  & $1.86$ \cite{De_Campos_Ferreira2024-jb} & $2.24\,(1.95)$ \cite{De_Campos_Ferreira2024-jb} & long  \\
dianion & $S_1^{2-}$ & $1.88$                  & $2.47\,(2.13)$ & short \\
        & $S_2^{2-}$ & $\sim 2.33$             & $2.79\,(2.73)$ & long  \\
\botrule
\end{tabular}

\end{table}

The pronounced charge-state dependence of the TERS intensity can be understood from the experimentally observed and TD-DFT-calculated excited-state energies, together with the corresponding transition dipole orientations summarized in Tab.~1. To reach a large resonant Raman enhancement for a molecule hanging on the tip, one of the molecular excitation energies has to be close to the laser energy. The transition dipole of the molecule involved in the resonant scattering must also be collinear with the nanocavity electric field, which is fulfilled for transition dipoles oriented along the long axis of PTCDA, due to the orientation of the molecule hanging on the tip. These conditions are met for the $D_{1}^{-}$ anionic excited state under 1.58 ~eV excitation and for the $D_{3}^{-}$ anionic excited state under 1.96~eV laser. There are no neutral or dianion excited states available, except for the $S_{1}^{2-}$ excited state of dianion at 1.88~eV, which is, however, oriented perpendicular to the nanocavity field (see Extended Data Fig.~\ref{figE6}). In contrast, excitation at 2.33~eV (Fig.~\ref{fig4}c) yields a more intense Raman signal for both the neutral and dianion states compared to the anion state, while preserving sharp intensity steps at charging voltages. This behavior can be explained by the near-resonance condition for dianion PTCDA, where the laser energy is, according to our measurements on MgO/Ag100 decoupling surface, close to $S_{2}^{2-} \to S_{0}^{2-}$ transition energy and for neutral PTCDA, where the laser energy is slightly below the $S_{1} \to S_{0}$ transition. Both these transitions have their respective transition dipoles oriented along the long axis of the molecule.

In addition, for the 1.58~eV excitation, we observed pronounced variations in the integrated TERS spectral intensity in the bias range of anion stability. The heatmap reveals a photoluminescence emission from $D_{1}^{-}$ along with its vibronic sidebands linearly shifting across a 1000--2000~cm$^{-1}$ range between $V^{+}$ and $V^{-}$ due to Stark shift of -24~meV/V, and crossing the nearly constant-energy Raman peaks. We attribute the increase in intensity at the intersection points to the previously described resonant Raman enhancement effect \cite{Jaculbia2020-tz}. The intermittent character of the resonances gives rise to the overall spectrum intensity oscillations. Similarly, a weaker photoluminescence peak around 1.35~eV corresponding to $D_{1}^{-} \to D_{0}^{-}$ can also be observed with a 1.96~eV laser excitation (see Extended Data Fig.~\ref{figE2}) and a nanocavity plasmon tuned towards the near-infrared region (Extended Data Fig.~\ref{figE7}). In this case, the photoluminescence is spectrally separated from the Raman fingerprints . The observed Stark shift constant of 28~meV/V is consistent with previous STML observations of the same system by Friedrich et al.~\cite{Friedrich2024-zp}.

\section*{Conclusion}\label{sec3}

In summary, we demonstrated an efficient optical readout of a molecular quantum dot charge state and proved its applicability by quantitative mapping of the electrostatic potential of a surface adatom. The main photophysical mechanism behind detecting the charge transitions in the TERS spectra and partial photon yields was identified as the switching between the near-resonant and non-resonant TERS conditions. The large difference in excited state energies for PTCDA charge states allows us to use a broad range of laser energies to obtain a sufficient contrast on both $V^+$ and $V^-$  required for electrostatic potential mapping via TERS-SQDM. By employing a tunable laser source, it will be possible to unambiguously disentangle the TERS and TEPL signals and optimize the absorption or emission of a given charge state. We envisage using this method to map excited states of molecules working as QD not only on the tip, but also on thick insulator layers \cite{Fatayer2019-hx,Sellies2025-vc,Patera2025-yp}, where they can be gated and switched between several charge states.

\section*{Methods}\label{sec11}

\subsection*{Sample and tip preparation}

Experiments were performed in an STM/AFM microscope (Createc GmbH) operating at 6~K and in an ultrahigh vacuum (UHV) environment. The Ag(111) crystal was prepared by standard cycles of Ar$^{+}$ sputtering and annealing at 550$^\circ$C. Purified PTCDA molecules were deposited by thermal sublimation from a home-built tantalum Knudsen cell at 327$^\circ$C for 2--3 minutes onto a clean Ag(111) sample in the STM head.

The tips were made from a 25~$\mu$m thick Ag wire sharpened by a focused ion beam. Subsequently, they were glued on a qPlus (type S1.0) tuning fork and on a standard STM tip holder using a conductive epoxy (EPO-TEK H21D). The resonant frequency of the sensor with $\sim 440\ \mathrm{\mu m}$ long tip was 30.96 kHz and $Q=50000$. After introducing the tip in UHV, it was further sputtered head-on with Ar$^{+}$ to remove carbon contaminants. The tip sharpness and plasmon resonance were tuned by 10~V voltage pulses and controlled indentation and characterized by electroluminescence spectroscopy, typically taken at 2.5--3.5~V bias and 1~nA current.

Based on our statistics of tens of microtips, we found that atomically sharp Ag tips with a weak van der Waals interaction with the substrate are required to lift a stable PTCDA QD onto the tip, where a shallow stabilization potential arises from a fine balance between local Ag--O bonding and long-range van der Waals attraction \cite{knol2021stabilization}. This condition typically corresponds to a frequency shift of $\Delta f \approx -1$~Hz at a setpoint of 10~pA and 100~mV bias voltage using 50~pm oscillation amplitude. The sharpness of the microtip was verified by making 2--3~nm indentations into the Ag surface at zero bias. Tips producing radially symmetric Ag clusters after indentation were selected, since the apparent cluster shape reflects a convolution of the deposited cluster and the microtip geometry. Such sharp tips usually exhibited a plasmon resonance at a single energy, as shown in Extended Data Fig.~\ref{figE7}.

\subsection*{Stability of the QD under laser illumination}

When PTCDA was lifted under laser illumination at powers typically used for TERS-SQDM measurements, the molecule toppled from the apex onto the tip shaft in most lifting attempts. Extended Data Fig.~\ref{figE1}\textbf{b} shows a successful lifting attempt under very low laser power (approximately 1/5 of typical TERS-SQDM power). Therefore, in most experiments, we lifted the molecule with the laser switched off, as shown in Extended Data Fig.~\ref{figE1}\textbf{a}. Once a stable QD tip was obtained, we switched on the laser and focused on the junction with a relatively low power. Keeping the power low effectively prevents flipping or detachment of the molecule, yet it ensures a sufficiently intense scattering signal yield. The tip height was finely adjusted to compensate for the laser-induced thermal expansion of the tip and the sample, thereby maintaining $V^{+}$ at the level corresponding to the unilluminated scenario.

\subsection*{TERS measurement}

We used a confocal optical setup described in our previous studies~\cite{De_Campos_Ferreira2024-jb,Dolezal2024-so}. The photon counts were measured using a single photon avalanche photodiode from Perkin-Elmer SPCM-AQR-15 referred to as APD1 in Figs.~\ref{fig3} and Extended Data Figs.~\ref{figE1} and \ref{figE4} , Micro Photon Devices PDM Series-100 referred to as APD2 in Figs.~\ref{fig1} and \ref{fig2} and Extended Data Figs.~\ref{figE5}, and \ref{figE5_5}, and the Andor Kymera 328i spectrograph equipped with Newton 920 CCD detector. The power of the laser polarized along the axis of the tip was typically kept between 500~nW and 10~$\mu$W, depending on the spectral overlap of the nanocavity plasmon resonance, the laser energy, and the Raman fingerprint region. The effective optical coupling of the incident laser field to the nanocavity can be estimated from the illumination-induced decrease of the first field emission resonance peak at 4.1~V~\cite{Liu2018-ek} (measured as demodulated current with closed feedback loop), which was, for 1-10~$\mu$W, only around 1\% at a setpoint of 400~pA and 50~mV lock-in modulation amplitude. Under these conditions, we typically detected a TERS signal from PTCDA of 50~kcps with APD1 at a bias voltage of +2~V and a tip-sample distance of 2.7~nm. The total photon collection efficiency is estimated to be 6\% at 700~nm wavelength for APD1 and 4\% for APD2.

\subsection*{Electrostatic potential calculation}

We denote $V^*$ the representation of the surface electrostatic potential $\Phi_\mathrm{S}$ sensed by the QD at imaging plane. It reads
\begin{equation}
    V^*(\mathbf{r}) =
    \frac{ V^+(\mathbf{r}) - V^-(\mathbf{r})}{V_0^+ - V_0^-} V_0^- -V^-(\mathbf{r}) 
 \label{V_star}   ,
\end{equation}
where \(V_0^-\) and \(V_0^+\) are the charging thresholds on the bare Ag(111) surface, taken at the corners of the image frame.

The total dipole of the surface object is calculated as
\begin{equation}
    P_\mathrm{\perp}
    =
    \varepsilon_{0}
    \int_{x_{\min}}^{x_{\max}}
    \int_{y_{\min}}^{y_{\max}}
   V^*(x,y,z) \, \mathrm{d}x \, \mathrm{d}y ,
\end{equation}
where the integration is performed over a sufficiently large imaged surface area and is height independent. 

\subsection*{Point spread function calculation and potential deconvolution}

The surface electrostatic potential $\Phi_\mathrm{S} $ is related to $V^*$ in the form of convolution with the point spread function $\gamma^*$ normalized to unit sum \cite{Wagner2019-bb,wagner2019theory,Bolat2024-fr} as
\begin{equation}
    V^*=\gamma^**\Phi_{\mathrm S}.
\end{equation}
 We calculated $\gamma^*$ by solving the axisymmetric Poisson equation in cylindrical coordinates for a grounded sphere-on-sphere tip above a grounded planar sample. The tip comprised a spherical body with radius of curvature $R=50~\mathrm{nm}$, terminated by a hemispherical apex of radius $r=0.5~\mathrm{nm}$ \cite{Yang2023-ot}, which represents a more realistic geometry than a completely flat tip considered by Wagner et al. \cite{Wagner2019-bb}. A localized test charge was placed 0.1~nm above the sample, and the resulting potential was evaluated 0.7~nm below the tip apex, corresponding to the position of the molecular quantum dot \cite{Wagner2019-bb}.

The Poisson equation was discretized using radial and vertical grid spacings of 50~pm and 25 pm, respectively. The sample, tip surface, and outer simulation boundaries were held at zero potential, and axial symmetry was imposed at the tip axis. Kernels were calculated for tip heights from 2 to 4 nm in 41 steps. The radial response was converted into a rotationally symmetric $93\times93$-pixel kernel covering $15.5\times15.5~\text{nm}^2$ and normalized to unit sum.

For Fig.~\ref{fig3}, the kernel at $z_\mathrm{ts}=2.6~\text{nm}$ was interpolated to the experimental pixel spacing and used to reconstruct $\Phi_{\mathrm S}$ from $V^*$. After subtraction of a constant background determined from the image corners, the deconvolution was performed in Fourier space using the regularized inverse
\begin{equation}
\widetilde{\Phi}_{\mathrm S}(\mathbf{k})
=
\frac{
\overline{\widetilde{\gamma}^{*}(\mathbf{k})}
}{
\left|\widetilde{\gamma}^{*}(\mathbf{k})\right|^2+\lambda
}
\widetilde{V}^{*}(\mathbf{k}),
\end{equation}
where the tilde denotes the Fourier transform, the overline denotes complex
conjugation, and $\lambda$ is the regularization parameter. We used $\lambda=6\times10^{-4}$. The experimental map and the kernel were zero-padded to accommodate the full linear-convolution size and suppress circular wrap-around artifacts. No additional smoothing was applied to the reconstructed surface-potential map.

To compare the influence of the tip geometry on the deconvolved image of $\Phi_\mathrm{S}$, i.e. the lateral resolution of SQDM, we calculated PSF and $\Phi_\mathrm{S}$ in Extended Data Fig.~\ref{figE8}. We verified that the image charge calculation of PSF using $100000$ image charges for a point charge placed 0.1~nm above the sample surface used by Wagner et al. \cite{Wagner2019-bb} shown in Extended Data Fig.~\ref{figE8}\textbf{a-c} yields an equivalent PSF to the numerical solution of the Poisson equation for a plate-capacitor model. If we consider a realistic sphere-on-sphere tip defined in Fig.~\ref{fig3}\textbf{a}, decreasing the bigger radius of curvature $R$ broadens the PSF, but with almost negligible effect for $R>>z_\mathrm{ts}$. The increase of the smaller radius $r$  effectively increases the tip-sample separation, leading to the broadening of the PSF.

\subsection*{Modelling of the bias-dependent charging transitions}

To describe the bias-dependent TERS intensity and frequency-shift traces
around the charging thresholds \(V^{+}\) and \(V^{-}\), we adapted the
charging-step model introduced for SQDM by Green~\cite{Green2018-uv}.
The TERS intensity was assumed to follow the average charge-state occupation
of the PTCDA quantum dot,
\begin{equation}
    I_{\mathrm{APD}}(V)
    =
    \sum_q p_q(V) I_q(V) ,
\end{equation}
where \(p_q(V)\) is the probability of occupation of the charge state \(q\) and
\(I_q (V)\) is the corresponding TERS intensity.

In a small voltage window around the charging threshold, only two charge states
contribute, and the $I_q$ can be treated as a constant. The expression above, therefore, reduces to
\begin{equation}
    I_{\mathrm{APD}}(V)
    =
    C + I_0 B(V),
\end{equation}
where \(B(V)\) is the occupation probability of one of the two charge states, \(I_0\) is the TERS contrast between the two charge states, and \(C\) is a constant offset.

Charging occurs when a localized electronic level of the quantum dot is shifted
across the Fermi level of the tip \cite{kocic2015periodic}. The thermally broadened occupation step was written as
\begin{equation}
    f(V)
    =
    \frac{1}{1+\exp\left[
    \frac{\alpha e (V_0-V)}{k_{\mathrm{B}}T}
    \right]},
\end{equation}
where \(V_0\) is the charging voltage and \(\alpha\) is the lever arm converting
a bias change into an energy shift of the localized level,
\(\Delta E=\alpha e\,\Delta V\). Equivalently, when all broadenings are
expressed on the voltage axis, the thermal width is \(k_{\mathrm{B}}T/(\alpha e)\).

The finite lifetime of the localized electronic state was included by
convolution with a Lorentzian kernel on the voltage axis,
\begin{equation}
    L_{\Gamma}(V)
    =
    \frac{1}{\pi}
    \frac{\Gamma}{V^2+\Gamma^2},
\end{equation}
where \(2\Gamma\) is the Lorentzian full-width at half maximum expressed in
volts. The corresponding energy broadening is \(\Gamma_E=\alpha e \Gamma\).
The intrinsic broadened charging step is then
\begin{equation}
    g(V)
    =
    \left[f * L_{\Gamma}\right](V).
\end{equation}

The vertical oscillation of the qPlus sensor modulates the tip--molecule
distance and therefore the charging voltage. For small oscillation amplitudes,
the distance dependence of the charging voltage was linearized as
\begin{equation}
    V_{\mathrm{ch}}(z)
    \simeq
    V_0
    +
    \frac{\mathrm{d}V_{\mathrm{ch}}}{\mathrm{d}z}
    (z-z_0).
\end{equation}
Thus, for a harmonic tip trajectory with oscillation amplitude $A_z$
\begin{equation}
    z(\theta)=z_0+A_z\cos\theta ,
\end{equation}
the charging condition is modulated in voltage by
\begin{equation}
    A_V
    \simeq
    \left|
    \frac{\mathrm{d}V_{\mathrm{ch}}}{\mathrm{d}z}
    \right| A_z .
\end{equation}
This linear mapping is expected, for example, in the plate-capacitor
approximation, where a change in tip--sample separation produces a proportional
change of the bias required to align the localized level with the tip Fermi
level.
For the photon-intensity traces, the measured signal corresponds to the
time-averaged charge-state occupation during the qPlus oscillation. The finite-amplitude occupation step was therefore calculated as
\begin{equation}
    B_I(V)
    =
    \frac{1}{\pi}
    \int_0^\pi
    g(V-A_V\cos\theta)\,\mathrm{d}\theta .
\end{equation}
Equivalently, this is a convolution of the intrinsic step \(g(V)\) with the
normalized arcsine kernel
\begin{equation}
    K_{A}(V)
    =
    \begin{cases}
        \dfrac{1}{\pi\sqrt{A_V^2-V^2}},
        & |V|<A_V, \\[6pt]
        0,
        & |V|\geq A_V .
    \end{cases}
\end{equation}
Thus,
\begin{equation}
    B_I(V)
    =
    \left[g*K_A\right](V),
\end{equation}
and the TERS intensity was fitted as
\begin{equation}
    I_{\mathrm{APD}}(V)
    =
    C_I + I_0 B_I(V;V_0,\Gamma,A_V,T).
    \label{Iapd_model}
\end{equation}

The frequency-shift signal was treated separately. We assumed that the charging
transition produces a step-like change in the electrostatic tip--sample force,
\begin{equation}
    F_{\mathrm{QD}}(V,z)
    =
    F_{\mathrm{off}}(z)
    +
    \Delta F(z)\,
    g\!\left[V-V_{\mathrm{ch}}(z)\right],
\end{equation}
where $F_{\mathrm{off}}(z)$ is the force in the reference charge state,
$\Delta F(z)$ is the force difference between the two charge states, and
$V_{\mathrm{ch}}(z)$ is the distance-dependent charging threshold. The same
intrinsic broadened step $g(V)$ as used above describes the charge-state
occupation probability. 
In frequency-modulated AFM, the frequency shift is given by the cosine-weighted average of the tip--sample force over one oscillation cycle,
\begin{equation}
    \Delta f
    =
    -\frac{f_0}{\pi k A_z}
    \int_0^\pi
    F_{\mathrm{ts}}(z_0+A_z\cos\theta)
    \cos\theta\,\mathrm{d}\theta,
\end{equation}
where $f_0$ is the unperturbed sensor resonance frequency, $k$ is the sensor
stiffness, $A_z$ is the oscillation amplitude, $z_0$ is the mean tip--sample
distance, and $\theta$ is the oscillation phase.
After linearizing $V_{\mathrm{ch}}(z)$ according to equations above and assuming
that $\Delta F(z)$ varies negligibly over the oscillation amplitude, the
dimensionless charging-induced frequency-shift line shape can be expressed as
a convolution of the intrinsic step $g(V)$ with the cosine-weighted arcsine
kernel
\begin{equation}
    K_{\Delta f }(V)
    =
    \begin{cases}
        \dfrac{V}{\pi A_V\sqrt{A_V^2-V^2}},
        & |V|<A_V, \\[6pt]
        0,
        & |V|\geq A_V .
    \end{cases}
\end{equation}
where $A_V$ is the oscillation-induced modulation amplitude of the charging
threshold on the voltage axis. Thus,
\begin{equation}
    B_{\Delta f}(V)
    =
    \left[g*K_{\Delta f}\right](V),
\end{equation}
where $B_{\Delta f}(V)$ describes the dimensionless shape of the
charging-induced frequency-shift feature.
After absorbing the force contrast $\Delta F$, the prefactor
$f_0/(\pi k A_z)$, the sign convention, and instrumental proportionality
factors into a scale factor $S$, the measured frequency-shift trace was fitted
as
\begin{equation}
    \Delta f(V) =C_{\Delta f}  + S B_{\Delta f}(V;V_0,\Gamma,A_V,T),
    \label{df_model}
\end{equation}
where $C_{\Delta f}$ is a constant frequency-shift background and $S$ is the
amplitude of the charging-induced feature.
We first performed amplitude dependent fits of data shown in Extended Data Fig.~\ref{figE5} with fixed temperature to \(6~\mathrm{K}\) and independent fitting parameters Lorentizan width $\Gamma$, the oscillation broadening $A_V$,  and charging threshold $V_0$ and the offsets ($C_{\Delta f}$ and $C_I$) and amplitudes ($S$ and $I_0$). The obtained values were used for the fits in Fig.~\ref{fig2}, where the only free parameter was the charging voltage $V_0$.

\subsection*{Electrostatic potential mapping with bias feedback}
We used a Kelvin module in Nanonis SPM control software for a bias feedback to acquire constant-height images of $V^{-}$ and $V^{+}$. Individual photon counts recorded as a TTL signal from APD2 were fed into PicoQuant DNS~102 electronics. Because the stochastic nature of the individual photon-counting digital signal prevented stable feedback, we used time-averaged logarithmically scaled analog voltage output of the electronics, \(V_{\mathrm{out}} \propto \log(I_{\mathrm{APD}})\), as the feedback signal.
For the acquisition of \(V^{-}\), the feedback time constant was set to \(4~\mathrm{ms}\) and the proportional gain to \(-1.1~\mathrm{mV/V}\), with a \(430~\mathrm{mV}\) setpoint of the analog output corresponding to \(I_{\mathrm{APD}} = 3~\mathrm{kcps}\). For the acquisition of \(V^{+}\), the time constant was set to \(1.2~\mathrm{ms}\) and the proportional gain to \(5~\mathrm{mV/V}\), with a \(1.45~\mathrm{V}\) setpoint corresponding to \(I_{\mathrm{APD}} = 36~\mathrm{kcps}\).

\subsection*{Time-dependent density-functional-theory calculations}
The electronic transitions of gas-phase PTCDA$^{2-}$ were calculated using
linear-response time-dependent density-functional theory (TD-DFT), as
implemented in Gaussian~16 \cite{Frisch2016Gaussian}. Calculations employed the
range-separated hybrid $\omega$B97X-D functional \cite{Chai2008} and the
6-31G* basis set. The ground-state geometry of the closed-shell dianion was
optimized, and the vertical excitation energies and transition dipole moments
were evaluated at this geometry. The geometry of the lowest singlet excited
state, $S_{1}^{2-}$, was subsequently optimized to estimate its relaxed-state
emission energy. The energy of $S_{2}^{2-}$ was additionally evaluated at the
relaxed $S_{1}^{2-}$ geometry.

The calculated vertical excitation energies were 2.47 and 2.79~eV for
$S_{1}^{2-}$ and $S_{2}^{2-}$, respectively. At the relaxed
$S_{1}^{2-}$ geometry, the corresponding transition energies were 2.13 and
2.73~eV. The calculated transition dipole moments are oriented along the short
molecular axis for $S_{1}^{2-}$ and along the long molecular axis for
$S_{2}^{2-}$. The calculations describe an isolated molecule and therefore do
not include the effects of the substrate, tip or dielectric environment.

\bibliography{sn-bibliography}

\section*{Data availability}

The data supporting the findings of this study and the scripts used for data analysis, simulations, and plotting are available from the corresponding authors upon reasonable request.  
\section*{Acknowledgments}
We gratefully acknowledge Ruslan Temirov for the discussions that initiated this study and Christian Wagner for the explanation of the electrostatic potential calculation and for providing the point spread function kernel. We thank Alessandro Pioda from SPECS Zürich for providing a demo version of the Nanonis Kelvin controller, Jaromír Kopeček for FIB milling of the Ag tips, and Tomáš Neuman for sharing the results of the $\mathrm{PTCDA}^{2-}$ - TD-DFT calculations.

\section*{Author contributions}
J.D. and M.Š. conceived the experiment. J.D. and A.S. performed the measurements. J.D. analyzed the data and created the figures. J.D. and M.Š. wrote the manuscript with the help of all authors.

\section*{Funding}
J.D. acknowledges an IOCB postdoctoral fellowship.
\section*{Competing interest}
The authors declare no competing interests.

\clearpage

\FloatBarrier
\begin{appendices}

\section*{Extended Data Figures}\label{secA1}

\setcounter{figure}{0}
\renewcommand{\thefigure}{\arabic{figure}}
\renewcommand{\theHfigure}{ED\arabic{figure}}
\renewcommand{\figurename}{Extended Data Fig.}




\begin{figure*}[!h]
\centering
\includegraphics[width=0.7\textwidth]{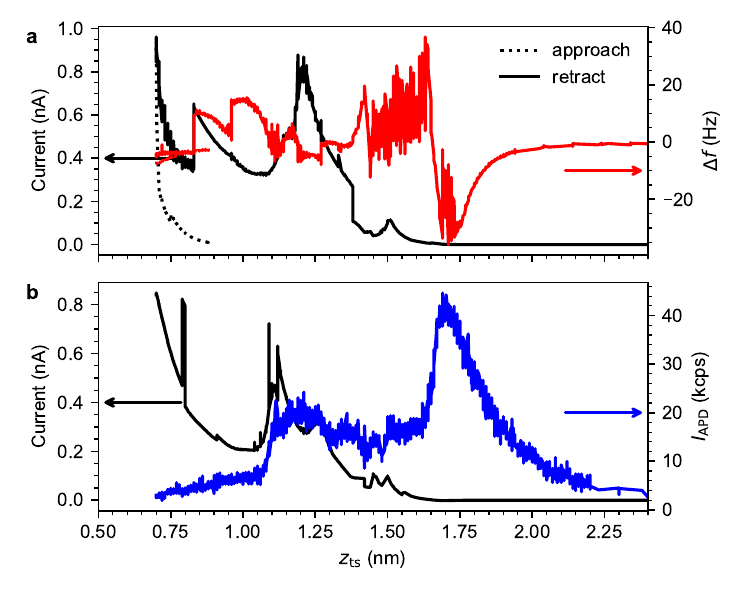}
\caption{
Lifting traces of PTCDA taken with \textbf{a,} Ag tip mounted on AFM sensor, without illumination. \textbf{b,} Ag tip in STM holder with a very low power illumination ($<$~500nW) with 1.96 eV laser. Sample bias was 1 mV and oscillation amplitude in \textbf{a} was 50 pm. The height where the molecule jumps to contact is $z_\mathrm{ts}=0.7$~nm.}
\label{figE1}
\end{figure*}

\begin{figure*}[!h]
\centering
\includegraphics[width=0.5\textwidth]{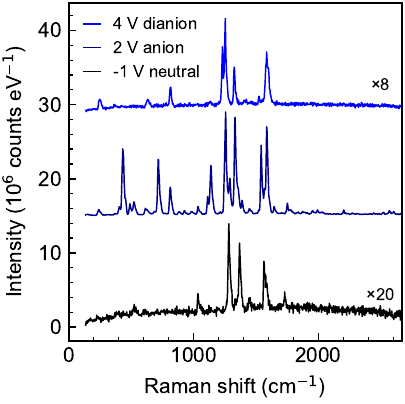}
\caption{
Raman spectra of 3 different charge states of the PTCDA quantum dot, taken at laser energy of 1.96~eV, $z_\mathrm{ts} = 2.7$ nm, $t = 20 $ s, $P = 10 \mathrm{\ \mu W}$.}
\label{figE3}
\end{figure*}

\begin{figure*}[t]
\centering
\includegraphics[width=0.5\textwidth]{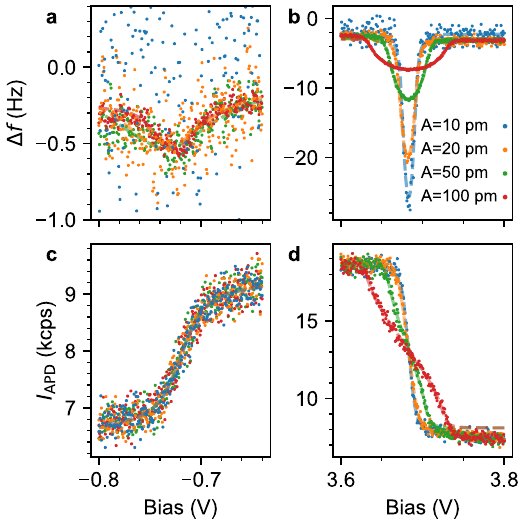}
\caption{
\textbf{a--d,} Bias-dependent frequency shift, $\Delta f(V)$, and intensity, $I(V)$, for different oscillation amplitudes around $V^{-}$ (\textbf{a}, \textbf{b}) and $V^{+}$ (\textbf{c}, \textbf{d}). Solid lines show fits using the equations (\ref{df_model}) and (\ref{Iapd_model}). Data were acquired at $z_\mathrm{ts} = 2.7$ tip height.}
\label{figE5}
\end{figure*}

\begin{figure*}[t]
\centering
\includegraphics[width=0.5\textwidth]{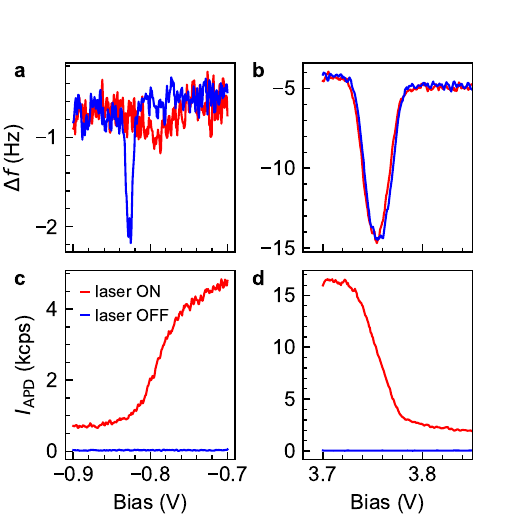}
\caption{
\textbf{a--d,} Bias-dependent frequency shift, $\Delta f(V)$, and APD intensity, $I_{\mathrm{APD}}(V)$, acquired with the 1.96-eV laser on (red, $P = 5~\mu\mathrm{W}$) and off (blue), using an oscillation amplitude of 20 pm at $z_\mathrm{ts} = 2.7$ nm height.}
\label{figE4}
\end{figure*}

\begin{figure*}[t]
\centering
\includegraphics[width=1\textwidth]{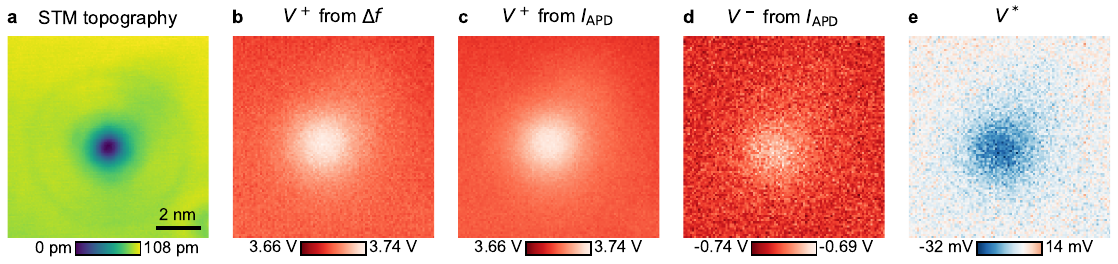}
\caption{\textbf{STM and electrostatic-potential imaging of a subsurface defect in Ag(111).}
\textbf{a,} STM topography of a subsurface defect in Ag(111) acquired in constant-current mode with an Ag tip at 100~mV and 10~pA.
\textbf{b,} $V^{+}$ map determined from the stack consecutively recorded $\Delta f$ images at varying biases around $V^{+}$.
\textbf{c,} $V^{+}$ map determined from the stack of consecutively recorded $I_\mathrm{APD} $ images at varying biases.
\textbf{d,} $V^{-}$ map determined from the stack consecutively recorded $I_\mathrm{APD} $ images at varying biases around $V^{-}$.
\textbf{e,} $V^*$, calculated from $V^{+}$ and $V^{-}$ using equation (\ref{V_star}). The dipole moment is -0.95~D. The tip--sample distance was $z_\mathrm{ts} = 2.7$~nm. Acquisition time was 193 s/image totaling $\approx 2h$ for 36 images used for the $V^{+}$ map and 2h and 12~min for 41 images used for the $V^{-}$ map.
}
\label{figE5_5}
\end{figure*}

\begin{figure*}[t]
\centering
\includegraphics[width=0.7\textwidth]{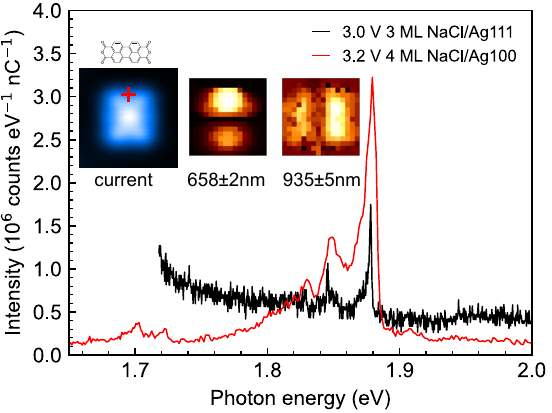}
\caption{
Electroluminescence spectra of PTCDA$^{2-}$ on NaCl/Ag acquired at the position marked by the red cross. The red spectrum was acquired for 5 s at 42 pA using a 150 gr./mm grating. The black spectrum was acquired for 180 s at 1.4 pA using a 1200 gr./mm grating. The inset shows a simultaneously recorded constant-height STM image and integrated electroluminescence maps of the dianion $S_{1}^{2-} \rightarrow S_{0}^{2-}$ transition $(658 \pm 2~\mathrm{nm})$ and the anion $D_{1}^{-} \rightarrow D_{0}^{-}$ transition $(935 \pm 5~\mathrm{nm})$, recorded on the same molecule as the red spectrum and revealing the orientation of their transition dipoles.}
\label{figE6}
\end{figure*}

\begin{figure*}[t]
\centering
\includegraphics[width=0.7\textwidth]{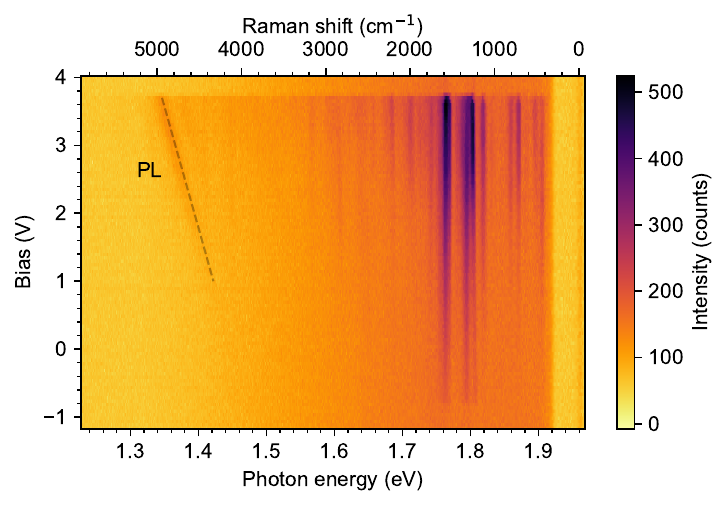}
\caption{
Heatmap of TERS spectra acquired with 1.96~eV laser as a function of bias voltage measured with 150 gr./mm grating. Stark-shifting of the photoluminescence peak of -28 meV/$\mathrm{V_{bias}}$ is marked with a dotted line.}
\label{figE2}
\end{figure*}

\begin{figure*}[t]
\centering
\includegraphics[width=0.7\textwidth]{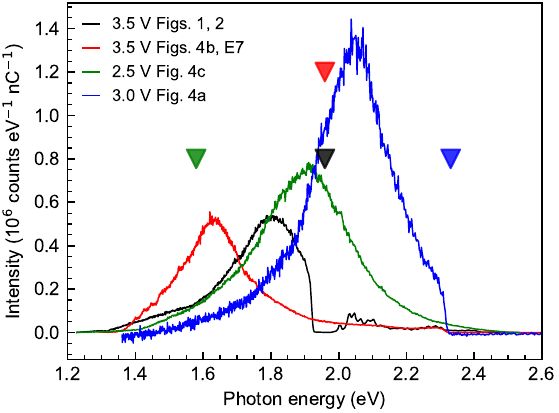}
\caption{
Spectral profile of electro-induced nanocavity plasmon resonances of the microtips used in our study, measured on the Ag(111) surface at the indicated bias voltages and nA currents. Corresponding laser energy used for TERS is marked with triangles. Notch and edge filters block the laser regions in the black and blue curves, respectively.}
\label{figE7}
\end{figure*}

\begin{figure*}[t]
\centering
\includegraphics[width=1\textwidth]{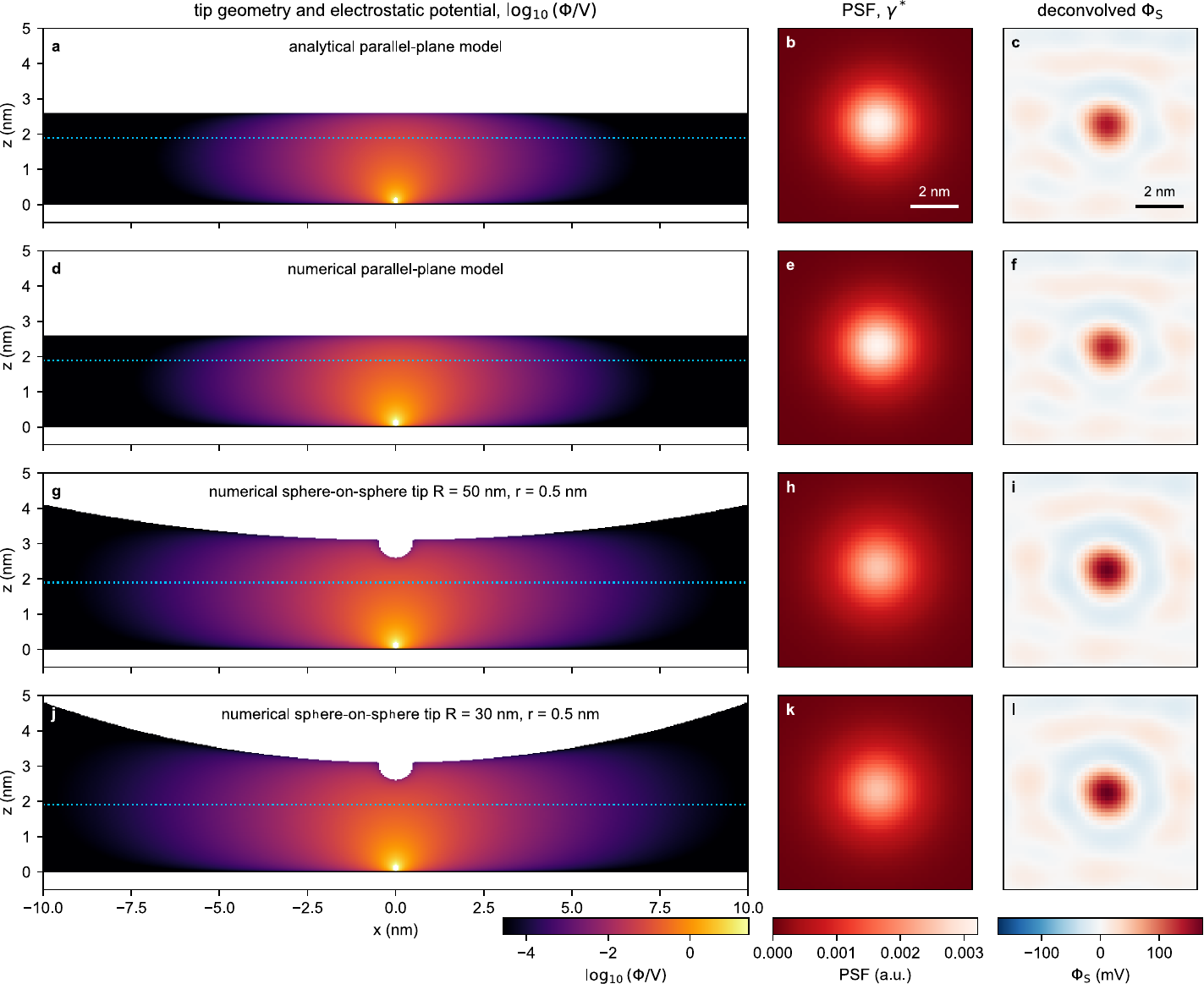}
\caption{\textbf{Comparison of electrostatic tip models used for SQDM deconvolution.}
\textbf{a--c,} Analytical parallel-plane model.
\textbf{d--f,} Numerical parallel-plane model.
\textbf{g--i,} Numerical sphere-on-sphere tip model with $R=50$~nm and $r=0.5$~nm.
\textbf{j--l,} Numerical sphere-on-sphere tip model with $R=30$~nm and $r=0.5$~nm.
For each model, the left panel shows the calculated electrostatic potential, $\log_{10}(\Phi/\mathrm{V})$, for a tip height of 2.6~nm; the grounded tip and sample are plotted with white color; the dotted line marks the PTCDA QD evaluation plane, located 0.7~nm below the tip apex. The middle panels show the corresponding normalized point spread functions, $\gamma^*$, and the right panels show the deconvolved surface potential $\Phi_\mathrm{S}$ of an Ag adatom on Ag(111) in Fig~\ref{fig3}. Identical color scales are used within each column.}
\label{figE8}
\end{figure*}

\end{appendices}

\end{document}